\documentclass[epsfig,graphics,twocolumn,floatfix,a4paper,showpacs]{revtex4}
\usepackage{amsmath,amsfonts,amssymb,graphicx,color,times,bbm,psfrag,dsfont}
\usepackage[english]{babel}
\usepackage[latin1]{inputenc}
\usepackage[T1]{fontenc}

\newcommand{\bed}{\[}
\newcommand{\eed}{\]}
\newcommand{\beq}{\begin{equation}}
\newcommand{\eeq}{\end{equation}}
\newcommand{\beqa}{\begin{eqnarray}}
\newcommand{\eeqa}{\end{eqnarray}}

\newcommand{\mean}[1]{\langle #1 \rangle}
\newcommand{\gras}[1]{\bold{#1}}

\newcommand{\tr}{\mathop{\mathrm{tr}}}

\usepackage{graphicx}
\usepackage[mathscr]{eucal}
\usepackage{amsmath}
\usepackage{amssymb}
\usepackage{epstopdf}
\usepackage{epsfig}
\def\one{\ensuremath{\hbox{$\mathrm I$\kern-.6em$\mathrm 1$}}}
\def\tr{ \mbox{tr}}

\begin{document} 

\title{Entropy and Exact Matrix Product Representation of the Laughlin Wave Function}

\author{S. Iblisdir$^1$,  J. I. Latorre$^1$ and R. Or\'us$^{1,2}$}

\affiliation{
${}^1$ Dept. d'Estructura i Constituents de la Mat\`eria,
Universitat de Barcelona, 647 Diagonal, 08028 Barcelona, Spain \\
${}^2$ School of Physical Sciences, University of Queensland, 
QLD 4072, Australia}
\date\today

\begin{abstract}

An analytical expression for the von Neumann entropy of the Laughlin wave function is obtained for any possible bipartition between the particles described by this wave function, for filling fraction $\nu=1$.  Also, for filling fraction $\nu=1/m$, where $m$ is an odd integer, an upper bound on this entropy is exhibited. These results yield a bound on the smallest possible size of the matrices for an exact representation of the Laughlin ansatz in terms of a matrix product state. An analytical matrix product state representation of this state is proposed in terms of representations of the Clifford algebra. For $\nu=1$, this representation is shown to be asymptotically optimal in the limit of a large number of particles. 

\end{abstract}
\pacs{03.75.Ss, 03.75.Lm, 03.75.Kk}
\maketitle



The puzzling behaviour of a cold gas of electrons confined in two spatial dimensions, and subjected to the effect of transverse magnetic fields, is the origin of two of the most studied phenomena in condensed matter physics: the integer and the fractional quantum Hall effects \cite{Klitzing80, Tsui82}. While it is possible to give a satisfactory explanation for the properties of the integer quantum Hall effect with filling fraction $\nu=1$, by means of transport and related models  \cite{transport}, a complete understanding of the fractional case is still missing. It is commonly believed that the interactions between the particles are essentially responsible for this situation. A relevant approach, in this respect, is the Laughlin ansatz for the wave function of the ground state of the system \cite{Laughlin}. While so far this state has only been proven to be an exact eigenstate of very specific Hamiltonians \cite{yoshi02} and for some specific values of the filling fraction, it contains the relevant properties that the ground state of the real system must have. The LWF has also been turned to be a valuable ansatz in describing the physics of rapidly rotating small atomic Bose-Einstein condensates \cite{pare01}. However, to check that the Laughlin wave function (LWF) is indeed the ground state of the fractional quantum Hall effect seems to be a difficult (computational) problem that has, so far, only been solved for systems made of a small number of particles, typically of the order of ten \cite{yoshi02}. 
												
In this paper, we wish to provide some clues on why this problem is so difficult, in the light of recent developments in the fields of quantum information science and many-body physics. It is now known that a profound relation exists between the difficulty to simulate numerically a quantum system in a given state and the von Neumann entropy between parts of the system this state exhibits, which can be understood in terms of the so-called Matrix Product States (MPS) \cite{mps}. With this relation in mind, we have computed the bipartite von Neumann entropy of a system as described by the Laughlin ansatz ($\nu = 1$). As we shall discuss, this computation is instructive in understanding the nature of the correlations exhibited by a system in such a state. To our knowledge, the behaviour of the von Neumann entropy was known only for the case of one particle versus the rest and in very specific situations \cite{ZZX02}. In addition, we will provide an MPS representation for this state, which is optimal in the thermodynamical limit. This representation will then be extended to some non-integer values of the filling fraction $\nu$. 

The LWF for $n$ particles and filling fraction $\nu$
reads \cite{Laughlin}
\begin{equation}
\label{eq:generallaughlin}
\Psi_\nu(z_1,\ldots,z_n)= \mathscr{N}_\nu(n)
\prod_{i<j}(z_i-z_j)^{1/\nu} 
\prod_{i=1}^{n} e^{-\vert z_i \vert^2/2} \ ,
\end{equation}
where $z_j, j=1,\ldots,n$ stands for the position of the particle $j$ in the $x-y$ plane written as a single complex coordinate $z_j=x_j+i y_j$. Note that, but in a few cases, the computation of the normalisation constant, $\mathscr{N}_\nu(n)$ seems to be a very difficult problem \cite{Itz}. 

It is convenient to introduce an orthonormal monoparticle basis defined as
\begin{equation}
\label{monoparticlebasis}
\phi_a(z_i)=\frac{1}{\sqrt{\pi a!}} \;  z_i^a \;  e^{-\vert z_i \vert^2/2} \qquad
a=0,\ldots,n-1.
\end{equation}

Using this basis, the LWF, for $\nu=1$, can be expressed as a Slater determinant:
\begin{equation}\label{eq:fermions}
\Psi_{1}(z_1,\ldots,z_n)=
\frac{1}{\sqrt{n!}}\sum_{a_1,\ldots,a_n=0}^{n-1}
\epsilon^{a_1\ldots a_n}\phi_{a_1}(z_1)\ldots \phi_{a_n}(z_n),
\end{equation}
where $\epsilon$ denotes the Levi-Civita completely antisymmetric 
tensor in $n$ dimensions, with the convention
$\epsilon^{0,1,\ldots,n-1}=-1$.

The above expressions show that it is possible to correctly describe the quantum state by means of a local Hilbert space of dimension $n$ for each particle. Since we have $n$ particles, the total dimension of the complete Hilbert space scales then as $n^n$. This superexponential scaling is at the heart of the difficulty to validate the ansatz (\ref{eq:generallaughlin}) by means of an exact diagonalisation 
\cite{yoshi02}. 

Let us now compute the von Neumann entropy for any subset of $k$ particles for a system of $n$ indistinguishable particles in the state $\Psi_{1}(z_1,\ldots,z_n)$. Note that this von Neumann entropy cannot be interpreted as the number of distillable EPR pairs. Due to the symmetrisation, it is impossible to associate a label with the particles and perform the appropriate distillation operations.

The reduced density matrix for a subset of $k$ particles out of $n$ reads
\begin{eqnarray}
\label{rho1}
\nonumber
&\rho_{k,n}\equiv\rho_{k,n}(w_1,\ldots,w_k;z_1,\ldots,z_k)=
\\ \nonumber &
\int dz_{k+1} \ldots dz_n \;
\Psi^*_{1}(w_1,\ldots,w_k,z_{k+1},\ldots,z_n) \times 
\\ & \Psi_{1}(z_1,\ldots,z_k,z_{k+1},\ldots,z_n).
\end{eqnarray}

From the orthonormality of the monoparticle basis, we get 
\begin{eqnarray}
\label{rho2}
\nonumber
\rho_{k,n}=&\frac{1}{n!}
\sum_{a,b,c}\epsilon^{a_1\ldots a_k c_{k+1}\ldots c_n}
\epsilon^{b_1\ldots b_k c_{k+1}\ldots c_n} \times
\\ & \phi^*_{a_1}(w_1)\phi_{b_1}(z_1)\ldots \phi^*_{a_k}(w_k)\phi_{b_k}(z_k).
\end{eqnarray}

This density matrix can be written in a diagonal form by introducing the following  
 set of orthonormal basis for the subset
of k particles:
\begin{equation}
\label{subsetbasis}
\Phi_{c}(z)\equiv
\frac{1}{\sqrt{k!}}\epsilon^{a_1\ldots a_k c_{k+1}\ldots c_n}
\phi_{a_1}(z_1)\ldots \phi_{a_k}(z_k),
\end{equation}
where $(z) \equiv (z_1,\ldots, z_k)$ and where the indices are sorted such that $c_{k+1}<\ldots <c_n$ and,thus, the combined index $c$ ranges from $1$ to $\left({n\atop k}\right)$.
It is then possible to see that 
all eigenvalues of $\rho_{k,n}$ are identical, that is
\begin{equation}
\label{finalrho}
\rho_{k,n}=\frac{1}{\left(n\atop k\right)}
\sum_{c=1}^{\left(n\atop k\right)}\Phi^*_c(w)\Phi_c(z).
\end{equation}
The von Neumann entropy then reads
\begin{equation}
\label{entropy}
S_{k,n}\equiv -{\rm Tr} \left(\rho_{k,n}\log_2 
\rho_{k,n}\right)=\log_2\left(n\atop k\right), 
\end{equation}
and gets its maximum value for a bipartition
of the system into two pieces of equal number of particles (this can be easily seen using strong subadditivity):
\begin{equation}
\label{asymptoticentropy}
S_{k,n}\le S_{\frac{n}{2},n}\sim n- \frac{1}{2} \log_2\frac{n\pi}{2}.
\end{equation}
Eq.(\ref{asymptoticentropy}) shows that the effective dimension of
the Hilbert space for half a Laughlin gas is  $O(2^n)$, which is a weaker scaling than 
the naive growth, $O(n^{n/2})$, of the Hilbert space for $n/2$ particles.

Let us now perform a similiar computation in the case $\nu=1/m$, when $m=2s+1$ is an odd positive integer different from $1$. As explained in Ref.\cite{Itz}, $\Psi_{\nu}(z_1,\ldots,z_n)$ can be expanded in terms of mutually orthogonal Slater determinants:
\bed
\Psi_{\nu}(z_1 \ldots z_{n})=
\prod_{i=1}^{n} e^{-\vert z_i \vert^2/2} \times
\eed
\beq\label{eq:slaterexpansion}
 \sum_{l_1, \ldots, l_{n}} g^{(s)}_{l_1 \ldots l_{n}}
\left|
\begin{array}{ccc}
z_1^{l_1} & \ldots & z_1^{l_{n}} \\
\vdots & \vdots & \vdots \\
z_{n}^{l_1} & \ldots & z_{n}^{l_{n}}
\end{array}
\right|,
\eeq
where the indices $l_1, \ldots, l_{n}$ are constrained by $ 0 \leq l_1 < \ldots < l_{n} \leq (2s+1)(n-1).$

Let $A(n,s)$ denote the number of coefficients in this expansion. It is easy to see that the rank $\tau_k$ of the reduced density matrix of any subset of $k$ particles is bounded as
\beq
\tau_k \leq A(n,s) \binom{n}{k}.
\eeq
Indeed, as we have seen in the study of the case $\nu=1$, $\binom{n}{k}$ is the Schmidt rank of each of the individual Slater determinants, and the Schmidt rank of a sum of kets  is smaller than or equal to the sum of the ranks of individual kets.  As abundantly discussed in Ref.\cite{Itz}, to give a closed form for $A(n,s)$ seems to be a very hard problem. Fortunately, an upper bound on $A(n,s)$ useful for our purposes can be derived as follows. Due to the constraints between the indices, the expansion 
(\ref{eq:slaterexpansion}) doesn't feature more than $((2s+1)(n-1)+1)^n$ terms. This trivial bound could be reached, would each index be allowed to range from $0$ to $(2s+1)(n-1)$ independently from the values taken by the other indices. But one can further constraint $A(n,s)$: since a given permuation of the columns of a determinant produces the same determinant (up to a sign factor), we have:
\beq
A(n,s) \leq ((2s+1)(n-1)+1)^n/n!
\eeq
Therefore, for $k=n/2$ and in the large $n$ limit, 
\beq
\tau_{n/2} \leq \frac{(2s+1)^n n^n}{n^n} 2^n=(4s+2)^n,
\eeq
and the von Neumann entropy obeys  
\beq\label{eq:vnentropynu}
S_{\frac{n}{2},n} \leq O(n \log(2s+1)).
\eeq
This bound can actually be slightly improved upon using the fact that all indices $l_0, \ldots, l_{n-1}$ assume different values

The dimensions of the matrices in an MPS 
representation of a given quantum state are related to the
von Neumann entropy of bipartitions  of the system when its degrees of freedom are
ordered on a line \cite{mps}. According to our previous calculation, an MPS representation of the LWF in terms of matrices of size $O(2^n)$ should be possible. More precisely, it should be possible to find a representation of the coefficients of the LWF written in the monoparticle basis in terms of products of matrices: 
\beq
\Psi_{\nu}(z_1 \ldots z_{n})= 
\frac{1}{\sqrt{n!}} 
\sum_{a_1,\ldots,a_n=0}^{n-1}
{\rm tr}\left( A^{[1] a_1} \ldots A^{[n] a_{n}} \right)\ \times \nonumber
\eeq
\begin{equation}\label{mpsproblem}
\phi_{a_1}(z_1) \ldots \phi_{a_{n}}(z_{n}),
\end{equation}
where the matrix $A^{[i]a_i}$ is associated with the particle $i$ being in the monoparticle state $\phi_{a_i}$. These matrices have a size which is the same for all values of $i, a_i$: $\chi \times \chi$. We will now see that the properties of Clifford algebras are well suited in order to find these matrices. The Clifford algebra $Cl(0,n)$ is  defined by
\begin{equation}
\label{clifford}
\{\gamma^a,\gamma^b\}=2 \delta^{ab} \qquad a,b=0,\ldots,n-1 ,
\end{equation}
where each matrix $\gamma^a_{\alpha\beta}$ has indices $\alpha,\beta=1,\ldots,\chi$ (see for example \cite{peskin}). 

Let us start with the case where $n$ is even. The representation theory of the Clifford  algebra dictates that $\chi=2^{n/2}$.  The matrices $\gamma^a_{\alpha\beta}$ provide the following MPS construction:
\begin{equation}
\label{mpslaughlin}
\epsilon^{a_1\ldots a_{n}}= \frac{-1}{(2i)^{n/2}}{\rm tr}\left(
\gamma^{a_1}\ldots \gamma^{a_{n}}\gamma_5\right),
\end{equation}
where $\gamma_5\equiv (-i)^{n/2} \; \gamma^0\ldots \gamma^{n-1}$. This result emerges
from the basic trace properties of the matrices $\gamma_5$ and $\gamma^a$, $a=0,\ldots , n-1$, and shows that all the matrices $A^{[i] a_i}$ can be taken the same for all particles but one. For example: 
\beqa\label{mpsdetail}
A^{[i]a_i} &\equiv& \gamma^{a_i} \qquad i=1,\ldots, n-1, \nonumber \\
A^{[n]a_{n}} &\equiv& \gamma^{a_{n}}\gamma_5.
\eeqa
Our construction shows that, in spite of the symmetry of the system,
the MPS representation is not made of a set of identical matrices.
This property was also observed for W-states in \cite{W}. 

The MPS construction (\ref{mpsdetail}) is asymptotically optimal in the sense that the matrices have, in the limit of large $n$, the minimal size. To begin with, let us first compute how large should the matrices be in an open boundary MPS representation. The rank of the indices of the different matrices then corresponds to Schmidt numbers of bipartite decompositions of the quantum state \cite{mps}. Given a bipartition of the system into two contiguous pieces, and assuming that the $\chi$  eigenvalues of the corresponding reduced density matrices of the two subsystems are all equal to $1/\chi$, we get a relation between $\chi$ and the maximum possible von Neumann entropy for that bipartition: $S_{\textrm{max}}=\log_2 \chi$. In general, though, the eigenvalues may not be equal, and therefore
$ \chi\ge 2^S \qquad {\rm (open\ boundary\ MPS)}$, $S$ denoting the actual von Neumann entropy of the bipartition. This argument fails for periodic boundary MPS  since
correlations between two sets of a bipartition can flow across \emph{two} boundaries.
Yet, it is possible to take the periodic-boundary MPS as a product of two effective
matrices $L_{\alpha\beta} R_{\beta\alpha}$
({\sl e.g.} for a bipartition of the system
$L_{\alpha\beta}=A^{[1]}_{\alpha\alpha_1}A^{[2]}_{\alpha_1\alpha_2}
\dots A^{[n/2]}_{\alpha_{n/2-1}\beta}$ and an analogous
construction for the right part of the system represented by $R_{\beta\alpha}$).
A simple Schmidt decomposition of the $L_{\alpha\beta} R_{\beta\alpha}$
construction shows that the Schmidt rank will now run up to $\chi^2$. The same argument as for open boundary conditions now shows that
\begin{equation}
\label{closedchiS2}
\chi\ge 2^{S/2} \qquad {\rm periodic\ boundary\ MPS} .
\end{equation}

The above considerations confirm that, in the case of periodic boundary conditions,
it is equivalent to argue that the effective rank of a bipartition has contributions coming from the two borders. As we have already seen, in our case the maximum possible entropy over all bipartitions corresponds to $S_{\frac{n}{2},n}\sim n$, which in turn implies that the dimension of the matrices of the Clifford algebra precisely matches the lower bound provided by this entropy for large $n$
\begin{equation}
\label{optimality}
\chi  =  2^{\frac{1}{2}S_{\frac{n}{2}, n} }  \sim 2^{ n/2 }.
\end{equation}
Therefore, the representation (\ref{mpsdetail}) is optimal in the limit of large $n$ values.

Explicit representation of the MPS construction can be obtained from the chain of isomorphisms 
\beq\label{eq:cliffordiso}
Cl(0,n+2)\approx Cl(0,2)\otimes Cl(0,n). 
\eeq
For $n=2$, set $\gamma^0=\sigma^x$, $\gamma^1=\sigma^y$ and $\gamma_5=\sigma^z$. A representation for $n+2$ is constructed from a representation of $n$ as follows:
\beqa
\gamma_{(n+2)}^i &=& \gras{1} \otimes \gamma_{(n)}^i, \hspace{0.2cm} i=0,\ldots, n-1, \nonumber \\
\gamma_{(n+2)}^i &=& \gamma_{(2)}^{a-n} \otimes \gamma_{5,(n)}, \hspace{0.2cm} i=n,n+1. 
\eeqa

A representation of the Clifford algebra for the case $n$ odd can be simply derived from a representation for $n-1$, upon taking
\beqa
\gamma_{(n)}^i &=& \gamma_{(n-1)}^i, \hspace{0.2cm} i=0,\ldots,n-2, \nonumber \\
\gamma_{(n)}^{n-1} &=& \gamma_{5(n-1)}.
\eeqa
Then, it can be shown that $\gamma_{5(n)}=(-)^n i^{3/2}$ and that the antisymmetric tensor is given by
\beq
\epsilon^{a_1\ldots a_{n}}= \frac{- i^{1/2}}{(2i)^{(n-1)/2}} {\rm tr}\left(
\gamma^{a_1}\ldots \gamma^{a_{n}} \right).
\eeq

It is also possible to extend the MPS construction of the Laughlin
wave function we have obtained for the case 
$\nu=1$ to the fractional case whenever $\nu=\frac{1}{m}$, where $m$ is an
integer. The basic idea is to use the property of the product of traces 
${\rm Tr}(A^1 \ldots A^{n}) {\rm Tr}(B^1\ldots B^{n})=
{\rm Tr}((A^1\otimes B^1)\ldots (A^{n}\otimes B^{n}))$ in a recursive way.
Let us exemplify the construction in the case of $1/\nu=2$, (bosonic statistics). 

Up to a global prefactor and a set of particle-dependent factors, the problem
of finding a faithful MPS representation essentially reduces to representing
\begin{eqnarray}
\label{prodm=2}
{\rm tr}\left(\gamma^{a_1^{(1)}}\ldots\gamma^{a_n^{(1)}}\gamma_5\right)
{\rm tr}\left(\gamma^{a_1^{(2)}}\ldots\gamma^{a_n^{(2)}}\gamma_5\right) \times \nonumber \\
\phi_{a_1^{(1)}+a_1^{(2)}}(z_1)\ldots\phi_{a_n^{(1)}+a_n^{(2)}}(z_n).
\end{eqnarray}

The structure of the coefficients is then correctly represented
by
\begin{equation}
\label{mpsm=2}
{\rm Tr}\left(\Gamma^{[1] c_1}
\ldots \Gamma^{[n] c_n}\Gamma^5\right),
\end{equation}
where, for all $i$, 
\begin{equation}
\Gamma_{1/2}^{c} = \sum_{a_1+a_2=c} \left(a_1+a_2 \atop a_1 \right) \gamma^{a_1} \otimes \gamma^{a_2}, \quad
\Gamma_{1/2}^5 = \left(\gamma^5\right)^{\otimes 2}.
\label{setofequations}
\end{equation}
It follows that the dimensions of the combined matrices
produces an effective $\chi=2^n$. In general,
an MPS representation of the LWF with filling fraction
$\nu=1/m$ for integer $m$ can be defined in terms of the matrices
$\Gamma_{1/m}^5= \left(\gamma^5\right)^{\otimes m}$,
\beq\label{setofequations2}
\Gamma_{1/m}^{c} = \sum_{a_1+ \cdots + a_m =c} \frac{(a_1 + \cdots + a_m)!}{a_1 ! \cdots a_m !} \gamma^{a_1} \otimes \cdots \otimes \gamma^{a_m},
\eeq
yielding an effective $\chi = 2^{\frac{mn}{2}}$. This value of $\chi$ is far above what one could expect to be the minimum from Eq.(\ref{eq:vnentropynu}) so that a much more economical construction should exist. Indeed, in the case $n=2$, we have found the following construction. The matrices $\Gamma^{c}_{1/m}$ can be viewed as acting on the $m$-fold tensor product representation of SU$(2)$. Define
$\Upsilon^{c}_{1/m}=G^{\dagger} \Gamma^{c}_{1/m} G$, where $G$ is a projector onto the spin-$m/2$ irreducible subspace contained in $(1/2)^{\otimes m}$ (a piece of Clebsch-Gordan matrix). After some algebra, we find that the explicit form of each matrix has
just a single entry 1 in its antidiagonal, $\Upsilon_{1/m}^c(c+1,n-c)=1$, and the rest is 0.
Further let $\Upsilon^{5}_{1/m}$ denote an $m+1 \times m+1$ matrix whose diagonal elements are defined as $(\Upsilon^{5}_{1/m})_{ii}=(-)^{i-1}\binom{m}{i-1}$ and whose off-diagonal elements are zero.
We have numerically checked $ \Upsilon^{c}_{1/m} $ and $\Upsilon^5$ yield an exact matrix product representation of $\Psi_{\nu}$ in the case $n=2$ for $m=1 \ldots 15$, and we believe that they also do for arbitrary values of $m$. Note that for $n=2$, the von Neumann entropy of the state of one particle can be easily calculated for any value of $\nu=1/m$. One gets $S_{1/m}=\sum_{j=0}^{m} \binom{m}{j} \log_2 \binom{m}{j}$. Clearly the size of the new matrices is now $m+1$ instead of $2^m$, that is exponentially more economical. We should note that  this representation has an interest as a starting point to get intuition on more involved cases. But, as such, it is not very useful in practice since the computation of a mean value $\mean{ V_1 \otimes V_2}$ requires an effort that scales as $O(m^2)$, which is the same as if using the Schmidt decomposition for $\Psi_{\nu}$.

For some filling fractions $\nu$, the foregoing analysis can be easily extended to study excited states of systems whose ground state is described by an LWF. Let us consider the simplest case of \emph{one} quasiparticle localised at a position $z_A$. The corresponding wave function reads \cite{pare01}
\beq
\Psi^{[z_A]}_\nu(z_1,\ldots,z_n) =\prod_{i=1}^n (z_i-z_A) \Psi_\nu(z_1,\ldots,z_n), 
\eeq 
Now since 
\bed
\prod_{i=1}^n (z_i-z_A)=\frac{\epsilon^{i_A, i_1 \ldots i_n} z_A^{i_A}  z_1^{i_1} \ldots z_n^{i_n}}
{\epsilon^{i_1 \ldots i_n} z_1^{i_1} \ldots z_n^{i_n}},
\eed
we have that $\Psi^{[z_A]}_{\nu'}(z_1,\ldots,z_n) =\epsilon^{i_A, i_1 \ldots i_n} z_A^{i_A} z_1^{i_1} \ldots z_n^{i_n}  \Psi_\nu(z_1,\ldots,z_n)$, with $\nu'=\nu/(\nu+1)$. With calculations akin to the ones performed above, an MPS representation of $\Psi^{[z_A]}_{\nu'}$ is easily derived. Defining $\Gamma^0(z_A)=\sum_{i_A=0}^n \gamma^{i_A} z_A^{i_A} \otimes \gras{1}$, we have (up to normalisation):
\beqa
\Psi^{[z_A]}_{\nu'}(z_1,\ldots,z_n) &=& 
\tr(\Gamma^0(z_A) \Gamma^{ a_1} \ldots \Gamma^{ a_n} \Gamma_5) \nonumber \\
& & \phi_{a_1}(z_1) \ldots \phi_{a_n}(z_n).
\eeqa
MPS representations of $m$-quasiparticle excited states can be computed likewise.

In summary, we have computed the von Neumann entropy of a Laughlin gas for any bipartition $(k,n-k)$ of the system under study and for various values of the filling fraction $\nu$. We have seen that this entropy grows at most linearly with $n$ and logarithmically with $\nu$. Since this quantity may be related to the difficulty of numerically simulate the gas, this computation sheds new light on why such systems are so difficult to study. Next, we have provided an MPS representation of the Laughlin wave function, that is asymptotically optimal in the case $\nu=1$. We believe this representation can be exploited to compute various quantities related to the LWF, such as its norm. This will be the subject of further investigation.

\emph{Acknowledgements.-}  We thank J.J. Garc\'ia-Ripoll for helpful discussions, L. Tagliacozzo for some preliminary numerical checks, and J.M. Escartin for a critical reading of this manuscript. This work has been supported by MEC (Spain), QAP (EU) and Grup consolidat (Generalitat de Catalunya).


\begin{thebibliography}{99}

\bibitem{Klitzing80}
K. von Klitzing, G. Dorda and M. Pepper, Phys. Rev. Lett. {\bf 45} 494 (1980).

\bibitem{Tsui82}
D. C. Tsui, H. L. Stormer and A. C. Gossard, Phys. Rev. Lett. {\bf 48} 1559 (1982).

\bibitem{transport}
R. B. Laughlin, Phys. Rev. B{\bf23} 5632 (1981); B. I. Halperin, Phys. Rev. B{\bf25} 2185 (1982); P. Streda, J. Kucera and A. H. MacDonald, Phys. Rev. Lett. {\bf 59} 1973 (1987); J. K. Jain and S. A. Kivelson, Phys. Rev. Lett. {\bf 60} 1542 (1988); M. B$\ddot{{\rm u}}$ttiker, Phys. Rev. Lett. {\bf 57} 1761; M. B$\ddot{{\rm u}}$ttiker, Phys. Rev. B{\bf 38} 9375; R. Landauer, IBM J. Res. Dev. {\bf 1} 223 (1957). 


\bibitem{Laughlin}
R. B. Laughlin, Phys. Rev. Lett. {\bf 50} 1395 (1983). 

\bibitem{yoshi02} D. Yoshioka, \emph{The quantum Hall Effect} (Springer-Verlag, Berlin 2002).

\bibitem{pare01}
B. Paredes, P. Fedichev, J.I. Cirac and P. Zoller, Phys. Rev. Lett. \textbf{87}, 10402 (2001). 

\bibitem{mps}
G. Vidal et al., Phys. Rev. Lett. {\bf 90} 227902 (1983);
S. R. White, Phys. Rev. Lett. {\bf 69} 2863 (1992); G. Vidal, Phys. Rev. Lett. {\bf 91} 147902 (2003); G. Vidal, Phys. Rev. Lett. {\bf 93} 040502 (2004); G. Vidal, cond-mat/0605597; F. Verstraete, A. Weichselbaum, U.  Schollw$\ddot{{\rm o}}$ck, J. I. Cirac and J. von Delft, cond-mat/0504305; U. Schollw$\ddot{{\rm o}}$ck, J. Phys. Soc. Jpn. {\bf 74} 246 (2005); F. Verstraete, D. Porras, J. I. Cirac, Phys. Rev. Lett. {\bf 93} 227205 (2004); F. Verstraete, J. J. Garc\'{\i}a-Ripoll and J. I. Cirac, Phys. Rev. Lett. {\bf 93} 207204 (2004); F. Verstraete and J. I. Cirac, Phys. Rev. B{\bf 73} 094423 (2006); D. P\'erez-Garc\' ia, F. Verstraete, M.M. Wolf and J.I. Cirac, quant-ph/0608197.

\bibitem{ZZX02}
B. Zeng, H. Zhai and Z. Xu, Phys. Rev. A{\bf 66}, 042324 (2002).

\bibitem{Itz}
P. Di Francesco, M. Gaudin, C. Itzykson and F. Lesage, Int. J. Mod. Phys. A{\bf 9} 4257 (1994). 


\bibitem{peskin}
M. E. Peskin and D. V. Schroeder, {\it An Introduction to Quantum Field Theory}, Addison-Wesley, 1995. 

\bibitem{W}
David Perez-Garc\' ia, private communication.



\end{thebibliography}
\end{document}